\documentclass[runningheads]{llncs}
\usepackage{graphicx}
\begin{document}
\title{Semi-Supervised Domain Generalization for Cardiac Magnetic Resonance Image Segmentation with High Quality Pseudo Labels}
%

\author{Wanqin MA\inst{1}\and
Huifeng Yao\inst{1} \and
Yiqun Lin\inst{1}\and
Jiarong Guo\inst{1} \and Xiaomeng Li\inst{1,2}
}
\authorrunning{F. Author et al.}
%
\institute{Department of Electronic and Computer Engineering, The Hong Kong University of Science and Technology 
\\
\email{\{wmaag,hyaoad,yiqun.lin,jguoaz\}@connect.ust.hk}
\and
The Hong Kong University of Science and Technology Shenzhen Research Institute
\email{eexmli@ust.hk}
}
\maketitle      
\begin{abstract}
Developing a deep learning method for medical segmentation tasks heavily relies on a large amount of labeled data. However, the annotations require professional knowledge and are limited in number. Recently, semi-supervised learning has demonstrated great potential in medical segmentation tasks. Most existing methods related to cardiac magnetic resonance images only focus on regular images with similar domains and high image quality. A semi-supervised domain generalization method was developed in~\cite{ref_lncs2}, which enhances the quality of pseudo labels on varied datasets. In this paper, we follow the strategy in~\cite{ref_lncs2} and present a domain generalization method for semi-supervised medical segmentation. Our main goal is to improve the quality of pseudo labels under extreme MRI Analysis with various domains. We perform Fourier transformation on input images to learn low-level statistics and cross-domain information. Then we feed the augmented images as input to the double cross pseudo supervision networks to calculate the variance among pseudo labels. We evaluate our method on the CMRxMotion dataset~\cite{ref_url1}. With only partially labeled data and without domain labels, our approach consistently generates accurate segmentation results of cardiac magnetic resonance images with different respiratory motions.
\\Code is available at: \url{https://github.com/MAWanqin2002/STACOM2022Ma}

\end{abstract}
\section{Introduction}
\begin{figure}
\includegraphics[width=\textwidth]{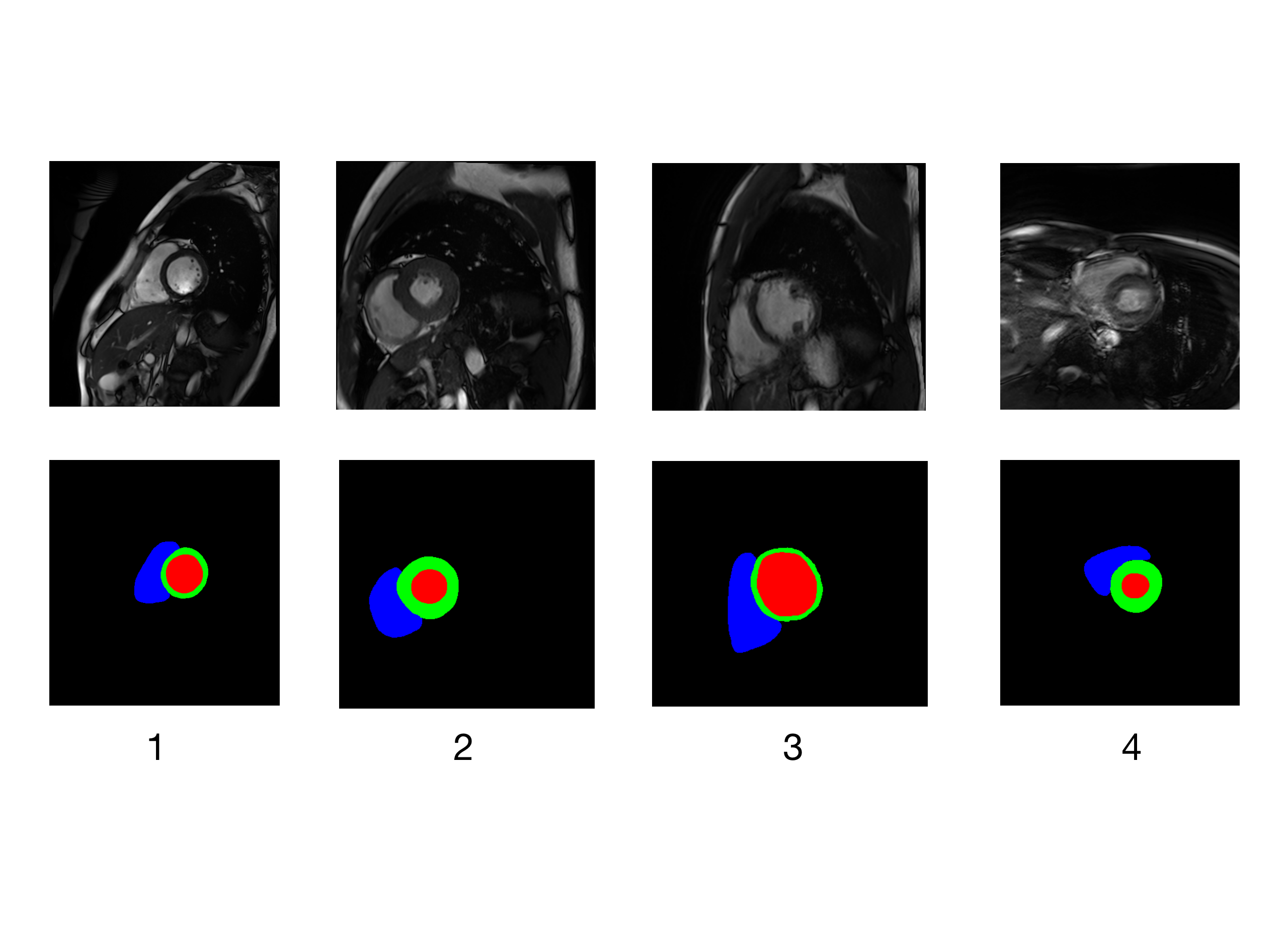}
\caption{These are representative 2D slices of images and masks from CMRxMotion Dataset~\cite{ref_url1}. The pictures are randomly selected from different patients with different stages. The original pictures are in the first row. The masks are in the second row. Labels 1,2,3,4 refer to different breath-hold stages: full breath-hold stage, half breath-hold stage, free breath stage, and intensive breath stage, respectively. 
There are three different colors in masks and refer to different structures: red refers to left ventricle blood pool, blue refers to the right ventricle blood pool, green refers to the left ventricular myocardium.} \label{fig1}
\end{figure}

In practice medical cases, accurate segmentation result is highly demanded as they can give important structural instructions on disease treatment and diagnosis. With the development of  many convolutional neural networks ~\cite{ref_lncs7,ref_article1,ref_article2,ref_article3,ref_lncs10,ref_lncs11,ref_lncs12} for medical image segmentation, deep learning methods are widely applied in varied medical tasks. Medical segmentation on cardiac magnetic resonance images is one of the most important issues. Most existing methods focus on pure cardiac segmentation in regular MRI with a similar domain. In practical cases, cardiac magnetic resonance (CMR) images are obtained from different patients with various equipments, provided under unstable imaging environments, affected by population shifts and unexpected human behaviors. This paper focuses on solving the issue raised in the CMRxMotion challenge~\cite{ref_url1}. The CMRxMotion dataset is collected from 45 patients with four different respiratory motion stages, which means the data is varied and leads to the problem of domain generalization. The example images of the CMRxMotion dataset are shown in Figure~\ref{fig1}.
However, the above methods are designed to solve medical segmentation tasks in regular CMR images and generate inferior results in varied CMR image domains. Hence, these methods are not suitable for solving problems raised in the CMRxMotion challenge~\cite{ref_url1}

To deal with the problems caused by domain generalization, many methods have been proposed. Li et.al~\cite{ref_lncs9} developed representative methods: skin lesion classification, and spinal cord gray matter segmentation. The method in~\cite{ref_lncs9} is aimed to learn consistent semantic information among multiple domains by capturing feature space. The above methods demand a fully-labeled dataset in each domain. However, in CMRxMotion dataset~\cite{ref_url1}, some images are unlabeled because of bad quality. Hence, the above method does not apply to CMRxMotion dataset~\cite{ref_url1}. Generating pseudo labels is an effective method to use the dataset, including unlabeled data fully. Chen et.al~\cite{ref_lncs6} introduced a cross pseudo supervision method by generating two segmentation networks. The two networks supervised each other by the corresponding pseudo labels. The unlabeled data in ~\cite{ref_lncs6} is from the same distribution as the labeled data; thus, pseudo labels from the same distribution can be used to improve another segmentation network directly. However, in CMRxMotion dataset~\cite{ref_url1}, the unlabeled images are from unknown distribution because of the variance of patients and different  respiratory motion, leading to a biased pseudo label.
\\
\\
Yao et.al~\cite{ref_lncs2} first attempt at combining Fourier transformation and cross pseudo supervision~\cite{ref_lncs6} to improve the quality of pseudo labels caused by domain generalization. For~\cite{ref_lncs2}, Fourier transformation is used to proceed with data augmentation and obtain important low-level statistics from the data, and a double network of cross pseudo supervision can train the model effectively by using pseudo labels. Specifically, it first randomly selects a sample to augment by Fourier transformation. Then, they use the original and augmented samples as input and pass them to the two cross pseudo supervision networks with the same structure. For the first network, they get prediction of augmented sample and the original sample. After that, they count the variance of the two predictions and generate the first one-hot vector. For the second network, the authors repeat the same steps as the first network and obtain the second one-hot vector. With two one-hot vectors, the two networks can supervise each other and improve the performance of the model.
\\
\\
Following this philosophy, in this work, we ensemble the semi-supervised framework consisting of Fourier transformation and cross pseudo supervision, which is proposed in~\cite{ref_lncs2}. The ensemble network is aimed to enhance the quality of pseudo labels, then can facilitate the whole training process. We conduct experiments on the Extreme Cardiac MRI Analysis Challenge under the Respiratory Motion dataset~\cite{ref_url1}. We find that, under the condition of extreme cardiac MRI Analysis, our method can always generate accurate segmentation results of cardiac magnetic resonance images with different respiratory motions.

\section{Methodology}
\subsection{Data Augmentation based on Fourier Transformation}
\begin{figure}
\includegraphics[width=\textwidth]{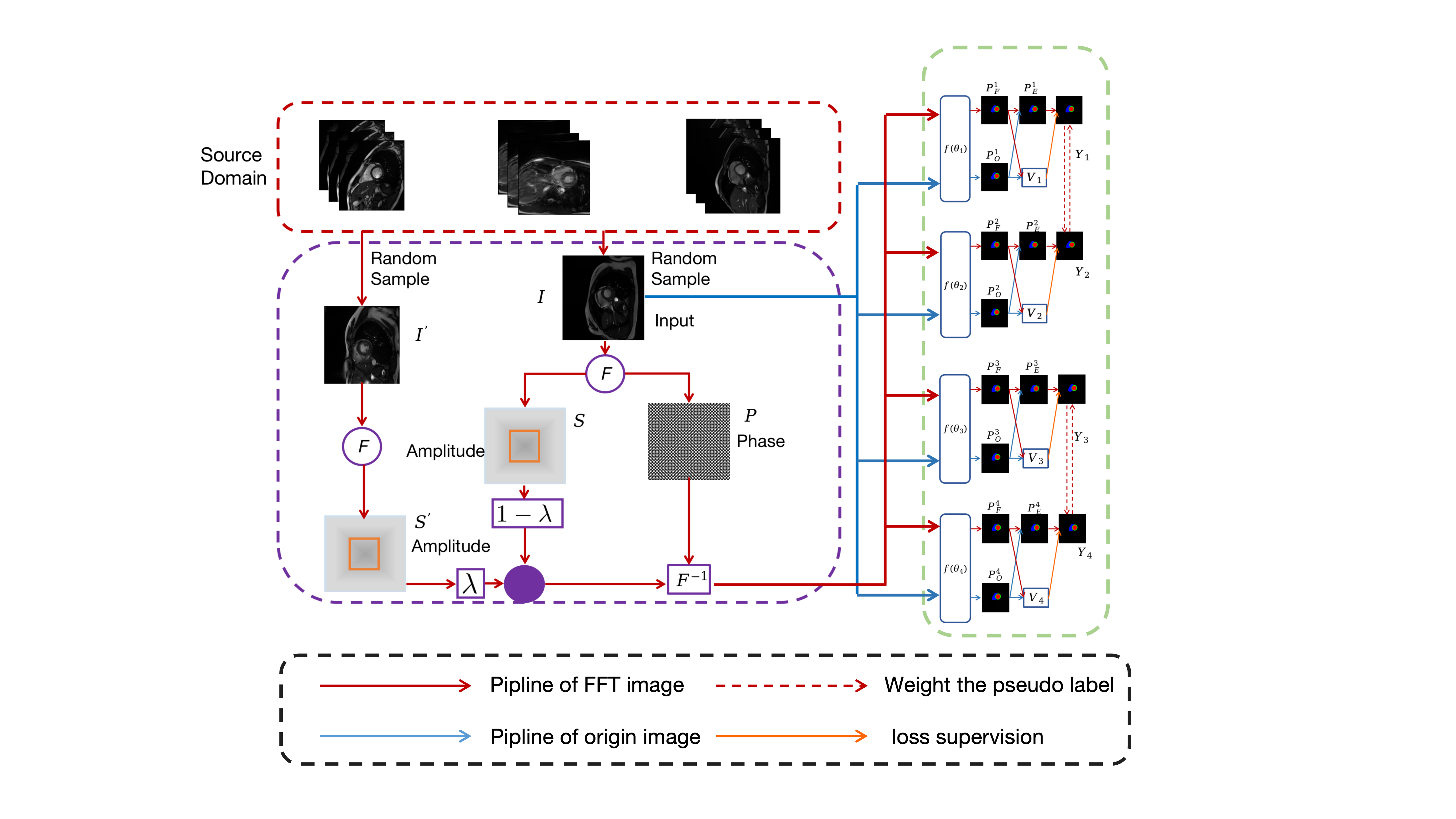}
\caption{The overall architecture of the proposed double confidence-aware cross-pseudo supervision network. The yellow area refers to the source domain. The orange area describes the steps of Fourier transformation. The green area is the network of the first CACPS model with DeepLabv3+ backbone.The blue area is the network of the second CACPS model with Resnet101 backbone.} \label{fig2}
\end{figure}

The overall methods are demonstrated in Figure~\ref{fig2}. During the training process, we input source images to the model without the score of the quality, the serial number of patient and  respiratory motion stage. Then we obtain an amplitude spectrum $S$ and a phase image $P$ by performing Fourier transformation $F$ on source images. Specifically, we can collect low-level statistics from amplitude spectrum and high-level semantics from phase images. After that, we repeat the steps above on another randomly selected sample image $I'$ and obtain a new amplitude spectrum $S'$.  With the information collected from the second image $I'$, we can augment the first image $I$ by:
\begin{equation}
I_{new} = (1-\lambda)S*(1-M)+\lambda S'*M\label{1}
\end{equation}

where $I_{new}$ is the augmented phase image; $\lambda$ is a parameter to control the proportion of the phase information of $I$ and $I'$; and $M$ is a binary mask to exchange the spatial range  of amplitude spectrum.To obtain low-frequency information, we set $M$ as the central region of the amplitude spectrum. Then we perform $F^{-1}$ inverse Fourier transformation on new sample to transform it from frequency domain to image domain. After that, we obtain the image sample $X$ which includes low-level information from another sample by Fourier transformation:
\begin{equation}
X=F^{-1}(I_{new},P)\label{2}
\end{equation}

\subsection{Ensemble of two Confidence-Aware Cross Pseudo Supervision}

In this section of the training process, we follow the strategy in ~\cite{ref_lncs2}, which introduced method Confidence-Aware Cross Pseudo Supervision (CACPS) to build the segmentation networks. We use double  Confidence-Aware Cross Pseudo Supervision with the same structure. We train two CACPS models separately.
\\In each CACPS, we build two parallel segmentation networks $f1$ and $f2$ (or $f3$ and $f4$ ) with the same structure but different initialized weights. Then, we take original image $I$ and the augmented image $X$ into the two networks above to obtain the predicted pseudo label of both images by:
\begin{equation}
P^{1(3)}_{F}=f(\theta _{1(3)})(X)
\end{equation}
\begin{equation}
P^{1(3)}_{O}=f(\theta _{1(3)})(I)
\end{equation}
\begin{equation}
P^{2(4)}_{F}=f(\theta _{2(4)})(X)
\end{equation}
\begin{equation}
P^{2(4)}_{O}=f(\theta _{2(4)})(I)  
\end{equation}
where $P^{1(3)}_{F}$ and $P^{2(4)}_{F}$ are pseudo labels of the augmented image; $P^{1(3)}_{O}$ and $P^{2(4)}_{O}$ are pseudo labels of the origin image; X is the augmented image, I is the origin image.
\\
\\
After that, we use the pseudo labels from one network to supervise the pseudo labels from another network because we do not use labels in supervised part to generate supervision signals for the unlabeled data. This method is from cross pseudo supervision~\cite{ref_lncs6}. The model cross pseudo supervision \cite{ref_lncs6} can generate pseudo label with high quality because of its consistent training data. However, the Extreme Cardiac MRI Analysis Challenge~\cite{ref_url1} data is from different domains with different levels of qualities. This varied dataset may cause low-quality prediction. Method Confidence-Aware Cross Pseudo Supervision~\cite{ref_lncs2} improves quality of pseudo label by counting variance between different input data. As showed in Figure~\ref{fig2}, we count the predictions from the original image and transformed image, $P^{1(3)}_{O}$ and $P^{1(3)}_{F}$. Then we calculate the average value $P^{1(3)}_{E}$ of $P^{1(3)}_{O}$ and $P^{1(3)}_{F}$ by: 
\begin{equation}
P^{1(3)}_{E} = (P^{1(3)}_{O}+P^{1(3)}_{F})/2
\end{equation}
Similarly, we obtain the $P^{2(4)}_{E}$ by: 
\begin{equation}
P^{2(4)}_{E} = (P^{2(4)}_{O}+P^{2(4)}_{F})/2
\end{equation}
To test the quality of the pseudo labels, we then calculate the variance $V_{1(3)}$ of predictions $P^{1(3)}_{O}$ and $P^{1(3)}_{F}$, the variance $V_{2(4)}$ of predictions $P^{2(4)}_{O}$ and $P^{2(4)}_{F}$:
\begin{equation}
V_{1(3)}=E[P^{1(3)}_{F}\log (\frac{P^{1(3)}_{F}}{P^{1(3)}_{O}})]
\end{equation}
\begin{equation}
V_{2(4)}=E[P^{2(4)}_{F}\log (\frac{P^{2(4)}_{F}}{P^{2(4)}_{O}})]
\end{equation}
where $E$ is the expectation value. If the variance value is large, this indicates the difference between two predictions is large and the quality of pseudo label is low.
Then, we obtain one-hot vectors $Y_{1(3)}$ and $Y_{2(4)}$ from the probability maps $P^{1(3)}_{E}$ and $P^{2(4)}_{E}$.
We define the confidence-aware loss function $L_{cacps} = L_{a}+L_{b}$ to improve the models by counting cross supervision loss:
\begin{equation}
L_{a}=E[e^{-V_{1(3)}}L_{ce}(P^{2(4)}_{E},Y_{1(3)})+V_{1(3)}]
\end{equation}
\begin{equation}
L_{b}=E[e^{-V_{2(4)}}L_{ce}(P^{1(3)}_{E},Y_{2(4)})+V_{2(4)}]
\end{equation}
where $L_{ce}$ is the cross-entropy loss.
We use dice loss as loss function in the supervised part. We formulate the supervision loss $L_{s}$.
\begin{equation}
L_s=E[L_{Dice}(P^{1(3)}_{O},G_{1(3)})+L_{Dice}(P^{2(3)}_{O},G_{2(4)}]
\end{equation}
where $L_{Dice}$ denotes the dice loss function and $G_{1(3)}(G_{2(4)})$ is the ground truth.
Hence, the whole loss function is:
\begin{equation}
L=L_s+\beta*L_{cacps}
\end{equation}
where $\beta$ is the CACPS weight. This parameter is help to keep balance between the two losses; $L_{cacps}$ is the confidence-aware loss.
We use the combination of two separated model's predictions as final results for the whole Condidence-Aware Cross Pseudo Supervision section.
\\
\\
As we actually use two CACPS models, in test procedure, we use the average of two Confidence-Aware Cross Pseudo Supervision Models' prediction as final result.

\section{Experiments and Results}
\subsection{Dataset and Preprocessing}
We evaluate our methods on a public medical image segmentation dataset from MICCAI 2022, the Extreme Cardiac MRI Analysis Challenge~\cite{ref_url1}. The dataset contains 360 3D clinical CMR scans from 45 healthy volunteers. A set contains 4 subsets obtained from a single volunteer under four levels of respiratory motion, including breath-hold state, halve the breath-hold state, breath freely states, and breathe intensively state. There are 2 CMR scans in a single subset. Among 360 CMR images, a set of 200 CMR scans are for training and validation, and the other 160 images are for the test. The annotations include three structures of the heart: left ventricle, left ventricle myocardium, and right ventricle. 
\\Regarding the preprocessing, we  generate 2D slices from the original 3D CMR images. Then we resize the slices of all 3D images to keep the correspondence in shape 512 x 512.

\subsection{Experiment Setting}
We ran the experiments using four Nvidia Geforce RTX 3090 GPUs with 96GB RAM on the Ubuntu20.04 system. Similar to~\cite{ref_lncs2}, we implemented the both CACPS models on PyTorch1.8. For the first CACPS model, we use DeepLabv3+~\cite{ref_lncs4} as the backbone. For the second one, we use the ResNet101~\cite{ref_url2} as backbone.

\begin{figure}
\includegraphics[width=\textwidth]{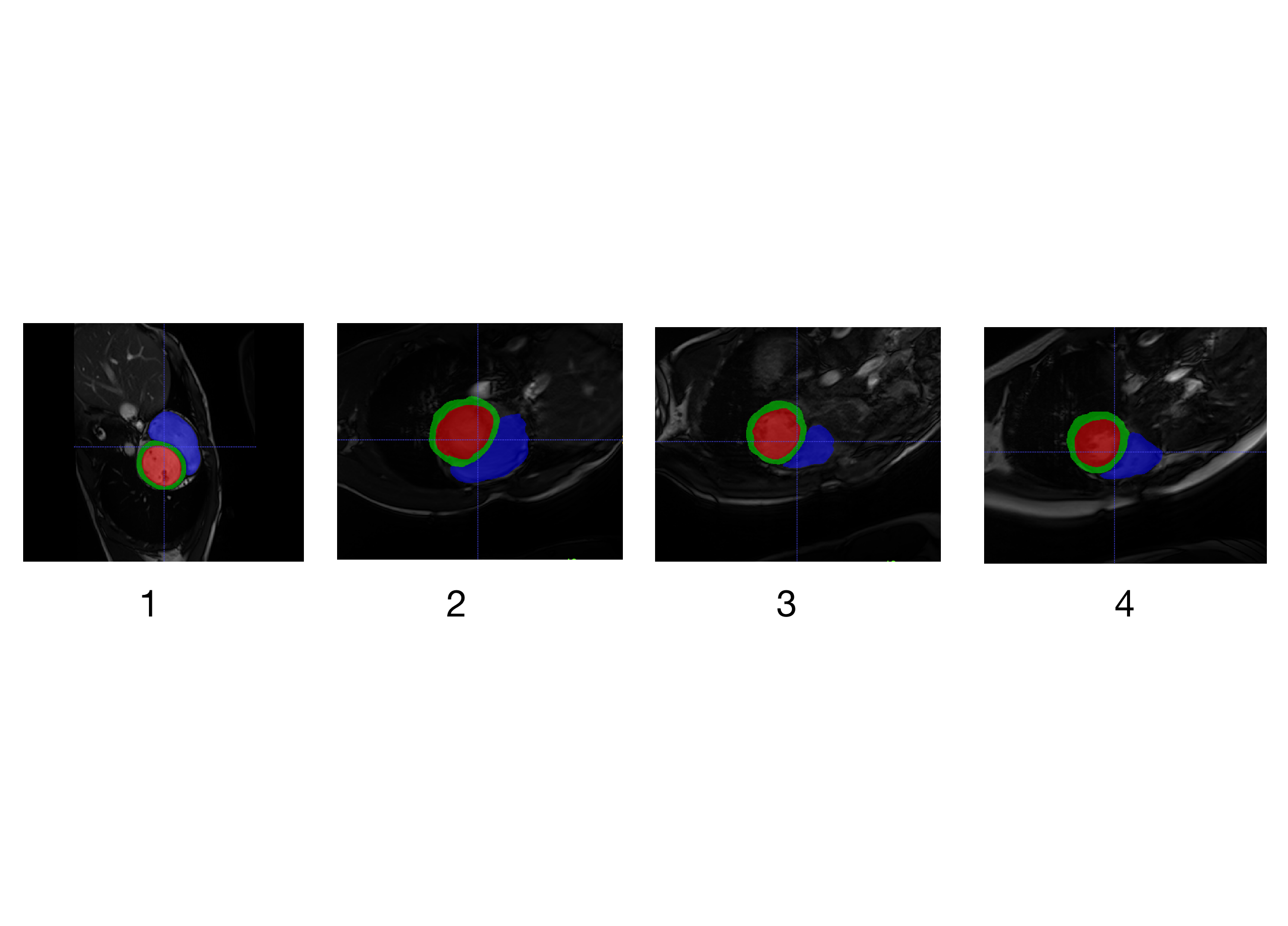}
\caption{These are representative segmentation results of Validation data. The pictures are randomly selected.  Labels 1,2,3,4 refer to different breath-hold stages: full breath-hold stage, half breath-hold stage, free breath stage, and intensive breath stage, respectively.  } \label{fig3}
\end{figure}

We can only access training and validation data during the challenge period. To train and test our model normally, we proceed following steps:
\\There are 160 CMR scans generated by 20 volunteers in total. We divide the training dataset into three parts according to its quality scores one, two or three. We name the group of images with quality score of one as domain 1. Similarly, we name domain 2 and domain 3. The ratio of the number of images in domain 1, domain 2, and domain 3 is 70:69:21. We notice that some images in the CMRxMotion Dataset is without masks. Hence, we use all training data to train the model through semi-supervised learning, and we select label data for validation. For Validation data, we use it to test our model. 
\\We choose AdamW as the optimizer of the network. The initial learning rate is 0.00001 and the number of epochs is 100. We use CosineAnnealingLR in pytorch to change learning rate effectively with cosine period in each epoch. And the batch size is 16. The weight of CACPS is 1.5. 

\subsection{Experiment Results}
We proceed with the experiment on three different methods. DeepLabv3+, a trained Confidence Aware Cross Pseudo Supervision model fine-tuning(Single CACPS(DeepLabv3+)) and the ensemble of double CACPS models(Double CAC
\\PS(DeepLabv3+Resnet101)). We use the dice scores given by CMRxMotion Challenge Website. The visualization of the segmentation results using our double CACPS models are showed in Figure~\ref{fig3}. 

\begin{table}
\caption{The table reveals the dices scores of all methods on validation dataset. LV-dice scores, MYO-dice scores, RV-dice scores refer to the accuracy of segmentation on left ventricle blood pool,the right ventricle blood pool, and the left ventricular myocardium, respectively.}\label{tab1}
\begin{tabular}{|l|l|l|l|l|}
\hline
Methods    &LV-dice scores     &MYO-dice scores   &RV-dice scores   &avg-dice scores\\
\hline
DeepLabv3+  & 0.799  & 0.060& 0.781 & 0.547\\
CACPS(DeepLabv3+)& 0.804 &0.552 & 0.802 &0.720\\
CACPS(DeepLabv3+Resnet101)&0.809 &0.594 & 0.810&0.738\\
\hline
\end{tabular}
\end{table}

Table~\ref{tab1} lists the average dice scores of all the methods. We find some meaningful information from this table: First, the use of the Confidence Aware Cross Pseudo Supervision network can enhance the quality of pseudo labels significantly. It shows that using cross pseudo supervision can get predictions with high quality. Second, we notice that combining the two CACPS networks  gains a considerable improvement in segmentation
performance among the state-of-the-art methods. It demonstrates that ensemble methods can improve the quality of pseudo labels more. Besides, it is noticeable that the performance of only DeepLabv3+ is not good. According to our analysis, the main reasons are as follows: First, some images in the CMRxMotion challenge are of bad quality under extreme respiratory motion. Although DeepLabv3+ is strong, it still can't learn data features effectively without augmentation. Second, we observe that the decline of the loss of model is prolonged after around 50 epochs when only using DeepLab3+. Without double CACPS networks, the learning efficiency of the model is low.

After the challenge, we receive the results of the average dice scores of double Confidence Aware Cross Pseudo Supervision network on the test dataset. The details of the result are list in Table~\ref{tab2} .
\begin{table}
\caption{The table reveals the dices scores of only CACPS(DeepLabv3+Resnet101) on test dataset. LV-dice scores, MYO-dice scores, RV-dice scores refer to the accuracy of segmentation on left ventricle blood pool,the right ventricle blood pool, and the left ventricular myocardium, respectively.}\label{tab2}
\begin{tabular}{|l|l|l|l|l|}
\hline
Methods    &LV-dice scores     &MYO-dice scores   &RV-dice scores   &avg-dice scores\\
\hline
CACPS(DeepLabv3+Resnet101)&0.866 &0.802 & 0.815&0.828\\
\hline
\end{tabular}
\end{table}

\section{Conclusion}
It's always very challenged to develop deep learning method for medical segmentation with limited annotation and various domain. Aiming at tackling this challenge, in this work, we develop a semi-supervised method to learn key information under domain generalization. Further, we use two strategies, Fourier transformation and cross pseudo supervision to improve the quality of prediction. Experiments on the Extreme Cardiac MRI Analysis Challenge under Respiratory Motion dataset~\cite{ref_url1} with partially labeled data reveals that when combining two confidence aware cross pseudo supervision. The resulting semi-supervised learning achieves accurate segmentation results on cardiac magnetic resonance images with different respiratory motions.

%
%
%
%

\end{document}